# Roaming Real-Time Applications
# - Mobility Services in IPv6 Networks

*Thomas C. Schmidt*[1 2], *Matthias Wählisch*[1]
{schmidt,mw}@fhtw-berlin.de

[1] Computer Centre, Fachhochschule für Technik und Wirtschaft Berlin,
Treskowallee 8, D-10318 Berlin, Germany

[2] FB Elektrotechnik und Informatik, HAW Hamburg, Berliner Tor 7,
D-20099 Hamburg, Germany

## Abstract

*Emerging mobility standards within the next generation Internet Protocol, IPv6, promise to continuously operate devices roaming between IP networks. Associated with the paradigm of ubiquitous computing and communication, network technology is on the spot to deliver voice and videoconferencing as a standard internet solution. However, current roaming procedures are too slow, to remain seamless for real-time applications. Multicast mobility still waits for a convincing design.*
*This paper investigates the temporal behaviour of mobile IPv6 with dedicated focus on topological impacts. Extending the hierarchical mobile IPv6 approach we suggest protocol improvements for a continuous handover, which may serve bidirectional multicast communication, as well. Along this line a multicast mobility concept is introduced as a service for clients and sources, as they are of dedicated importance in multipoint conferencing applications. The mechanisms introduced do not rely on assumptions of any specific multicast routing protocol in use.*

## 1. Introduction

The designation of the Internet has changed. Networked devices, formerly placed at scientist and business desks, are now consumer parts and serve for information, communication and entertainment. Their dedication follows trends as clearly lie in mobility and ubiquitous computing, at present. The vision of nomadic users at roaming devices performing synchronous communication, such as voice or videoconferencing over IP (VoIP/VCoIP), is around and raises new challenges for the Internet infrastructure.

This article is written from the perspective of real-time conferencing requirements within a mobile Internet. It confronts current mobility solutions with the principal communication characteristics under request. It further reports on ongoing work of analyzing, accelerating and smoothing mobility handover procedures, designed to operate transparently with respect to bi-directional multicast communication. Along this line we introduce and discuss M-HMIPv6, a new multicast handover scheme within a Hierarchical Mobile IPv6 environment.

In conferencing scenarios addressability raises the first major issues. To globally call a device, a routable IP address must be in use. On a large scale such address space is only provided by IPv6. To identify a communication partner's current device, a supplementary global user locating scheme is needed. Such mechanism is proposed in [1] or, alternatively, in a simpler ready-to-use approach in [2]. In addition, multicasting is needed to enable group conferencing without placing the burden of dedicated group-server infrastructure. All of the above functionalities have been implemented in videoconferencing systems, e.g. the DaViKo software [3],[4].

At the same time synchronous real-time applications s. a. VoIP and VCoIP place new demands on the quality of IP mobility services: Packet loss, delay and delay variation / jitter in a constant bit rate scenario need careful simultaneous control. A spoken syllable is about the payload of 100 ms continuous voice traffic. Each individual occurrence of packet loss above 1 %, latencies over 100 ms or jitter exceeding 50 ms will clearly alienate or even distract the user. For further aspects we refer the reader to [18].

Thus in real-time communication available techniques of hiding packet loss at the cost of delay and jitter or vice versa are of limited use. These requirements impose strong challenges on a mobile Internet scenario. Challenges are even tightened by multicast-based group communication. In conferencing scenarios each member commonly operates as receiver and as sender. Multicasting in a mobile Internet environment poses the fundamental problem that routing information bases need to be built from possibly rapidly changing group member locations. A mobile environment consequently needs to cope with the tardy source specific construction of multicast routing trees.

This paper addresses these central aspects of handover in real-time mobility. It is organised as follows. The current mobility state of the art is briefly recalled in section 2. In section 3 we analyse the local handover performance of MIPv6 over 802.11 and its potential for optimisation. Section 4 discusses the general problem of MIPv6 under real-time requirements. Section 5 presents our proposal for extending Hierarchical MIPv6 architecture with multicast mobility. Conclusions and an outlook follow in section 6.

## 2. Related Works

### 2.1. Internet Mobility

There are several attempts to cope with mobility under IPv6. The most fundamental approach is the Mobile IPv6 (MIPv6) Internet-Draft [5]. MIPv6 transparently operates address changes on the IP layer as a device moves from one network to the other by sustaining original IP addresses in a Home Address Destination Option and hiding the different routes to the socket layer. In this way, hosts are enabled to maintain transport and higher-layer connections when they change locations. An additional component, the MIPv6 Home Agent, preserves global addressability, while the mobile node is away from home.

An inventive idea to obtain mobility on the IP layer is built on the location independence of multicast addresses [6]: Equipping each mobile node with its individual multicast address, a correspondent node can send packets without knowledge on current location of the mobile. To preserve connection oriented transport ability the Home Address Option of MIPv6 is used. Handoff speed is justified by a vicinity argument. Besides security issues, the major drawback of the multicast based mobility is due to the asymmetry of multicast routing: Correspondent nodes cannot be mobile, themselves.

An alternate approach to application persistency under mobility is grounded on the Stream Control Transmission Protocol (SCTP) [9]. Initially designed for network redundancy, SCTP allows for multihoming of a single socket. The ‘Add IP’ proposition [10] of extending this functionality to adding and deleting IP addresses, gives rise to an address handover on the transport layer. Mobile SCTP [11] carries the justification of performing a rapid handover on the client side, only, without any provisions in the network infrastructure. Mobile SCTP, though, conflicts with single bound layer 2 protocols s. a. 802.11, connectionless flows, multicast traffic and does not support MSCTP node discovery. Thus there are strong arguments for gaining transport mobility as a combination of SCTP and MIPv6, instead.

As an application layer protocol SIP [1] provides some mobility management to session-based services on the basis of MIPv6. Employing the SIP server as an application specific home agent, handoff notifications are traded via regular SIP messages to the home server (`register`) and the correspondent node (`reinvite`). As SIP mobility operates above the transport layer, it inherits all underlying delays in addition to its own signalling efforts. SIP mobility thus comprises a foreseeable latency problem.

Our present work is concerned with mobility based on the Mobile IPv6, as the current draft elegantly provides full infrastructure support transparent to applications. MIPv6, on the one hand, conducts changes of networks instantaneously and independent of the subnetwork layer technology. Mobile IPv6 handovers, on the other hand, currently may cause an inaccessibility of nodes during seconds on top of layer 2 delays. Current attempts to improve handover performance rank around two ideas: A proxy architecture of Home Agents is introduced by the Hierarchical Mobile IP (HMIPv6) [12], whereas latency hiding by means of handover prediction operated at access routers is proposed by the Fast Mobile IP (FMIPv6) [13]. For a comparative analysis see section 4.2.

### 2.2. Multicast Mobility

There are three principal approaches to bring multicast into concordance with a mobile Internet (see [16] for a more detailed overview): The simplest one rigorously hides mobility to multicast routing by tunnelling all group communication via the Home Agent. This Bi-directional Tunnelling (BT) is advised by MIPv6 [5] as minimal multicast support. It is directly proportional to MN’s distance from home, how well BT is suited for multicast mobility management. This becomes apparent by observing that the additional triangular routing efforts, as well as the effort for handover Binding Updates are characterised by packet transport between the Home Agent and the Mobile Node.

The contrary approach to cope with multicast mobility shows all mobility operations to the multicast routing and is known as Remote Subscription (RS). By using link-local addresses, RS admits full - and thereby optimal - multicast routing features. However, for multicast listeners on handover, these routing topologies unfold unbearably slow. A mobile multicast source sending with its link-local address is immobile with respect to multicast application persistence, as multicast sessions are source address aware.

Source address awareness in the literature (s. [16] and references therein) is mainly discussed in the context of Source Specific Multicast (SSM). It is noticed that source addresses in multicast packets carry a dual meaning of source identifier and location information. This of course is true also in the unicast case and the loss of natural coincidence of source and location is what Mobile IP is about. The specific aspects of multicast mobility ground in routing and its source address orientation. The foremost sensible countermeasure on visible source address changeovers is to distinguish between applicative and topological source address on Internet protocol level, as Mobile IPv6 does. We will discuss such a procedure in section 5.

The third approach searches for a compromise in the form of intermediate nodes acting as agents: Multicast agents join a multicast group on behalf of a Mobile Node within the foreign network, thereby hiding part of MN's motion. Communication from the agent to the Mobile Node then usually continues through a tunnel. Source mobility at the micro-mobility scale is supported by tunnelling traffic to the multicast agent. Macro-mobile handovers, i.e. transfers from multicast agent to agent, conventionally are performed, causing session restarts due to change of source address and possible reconstruction of the routing tree.

Agent-based approaches may adapt well to the network of receivers and – while reducing rapid mobility – preserve the multicast nature of the traffic. The major drawback comes from a specific session support within a dedicated infrastructure formed from these agents. Further on in this paper we will discuss an integrative approach for a mobility infrastructure which seamlessly operates both, unicast and multicast mobility handovers.

## 3. Measuring local handover of MIPv6 over 802.11 Wireless

### 3.1. Object of Investigation

Timing of handover procedures in MIPv6 forms a critical issue in real-time scenarios. In entering a new IP network, i.e. after completing the layer 2 (L2) handoff, the Mobile Node (MN) instantaneously has to perform an automatic address reconfiguration followed by binding updates with its Home Agent (HA) and the Correspondent Node (CN). During this handover procedure the MN is unable to communicate until the HA has learned its new Care-of Address (CoA). Packets may then proceed through the HA as a forwarder with the likely result of increased delay and jitter.

The temporal performance under network changes is determined by two different mechanisms: The (link) local procedures of L2 handoff with succeeding readdressing on the one hand, which only depend on local subnet topology. The distant updates with HA and CN on the other hand are dominated by network geometry. As geometry effects are discussed in section 4 we first concentrate on the empirical performance analysis of local handover procedures, thereby trying to identify L2 as well as L3 latencies.

### 3.2. Experimental Setup

In our current experiments we focus on the local handoff procedures in MIPv6 over 802.11b wireless LAN following the setup of fig. 1. Starting from its home network a Mobile Node proceeds through two 802.11b wireless LANs. All subnets involved are directly attached to one router.

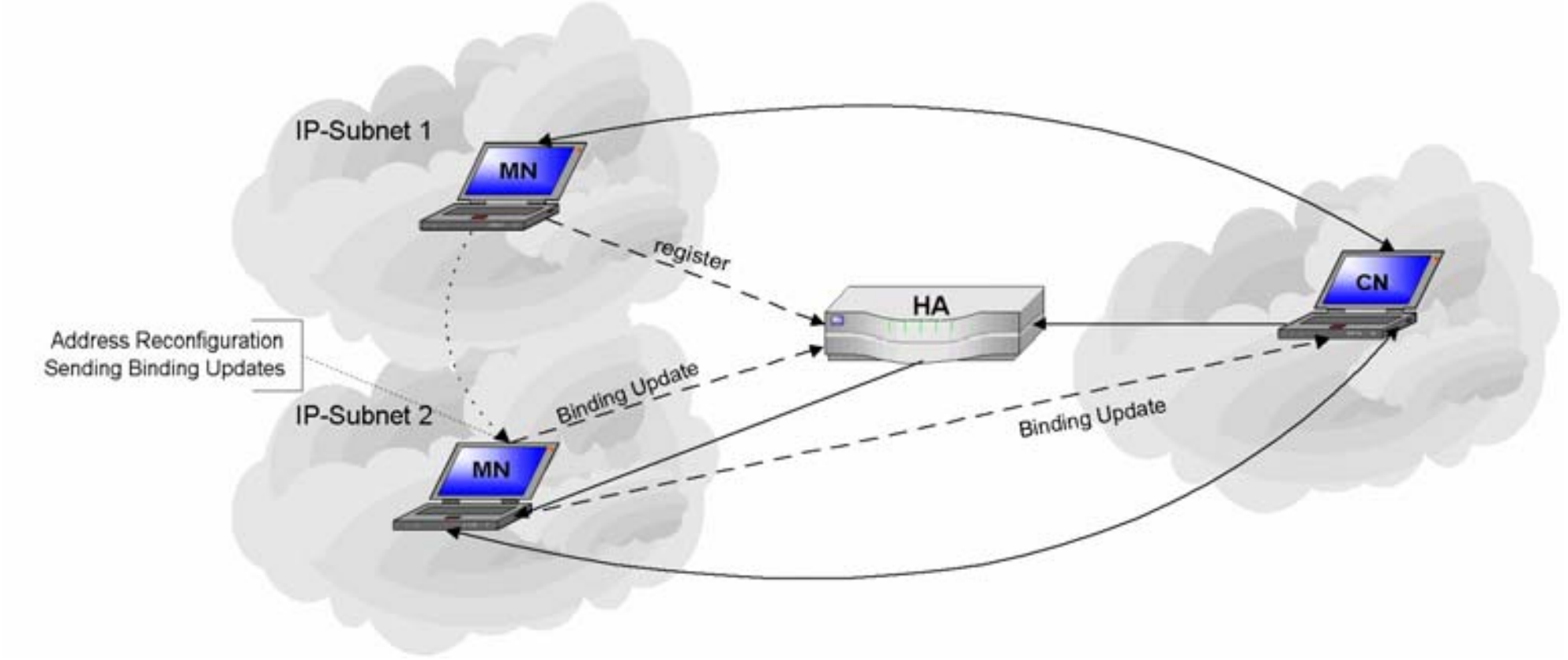

**Figure 1: Mobile Scenario**

Our test bed consists of a Debian Linux machine as a home agent, a router based on FreeBSD and Win2000 clients as well as Debian Linux for Mobile and Correspondent Nodes[i]. Stacks of the MNs and the router advertisement daemon were modified with respect to temporal and MIP-signalling (DAD-suppression) behaviour. Our WLAN is built of 802.11b access points "RoamAbout" of Enterasys Networks and corresponding NICs of the same manufacturer.

Experiments were performed using a simple test probe: The MNs send and receive numbered and time-stamped UDP packets following a predefined trigger of typically 10 to 20 ms. These packets are reflected by the CN. All events were packet wise recorded using a network sniffer.

### 3.3. Results

Packet loss, packet roundtrip time and jitter occurrences are shown in figure 2 as averages over events of the Linux MN's traversal. With a router advertisement interval (mininterval=37 ms, maxinterval=50 ms) of at most 50 ms 90 % of the events exhibit an interval of network disturbance below or equal to 100 ms.

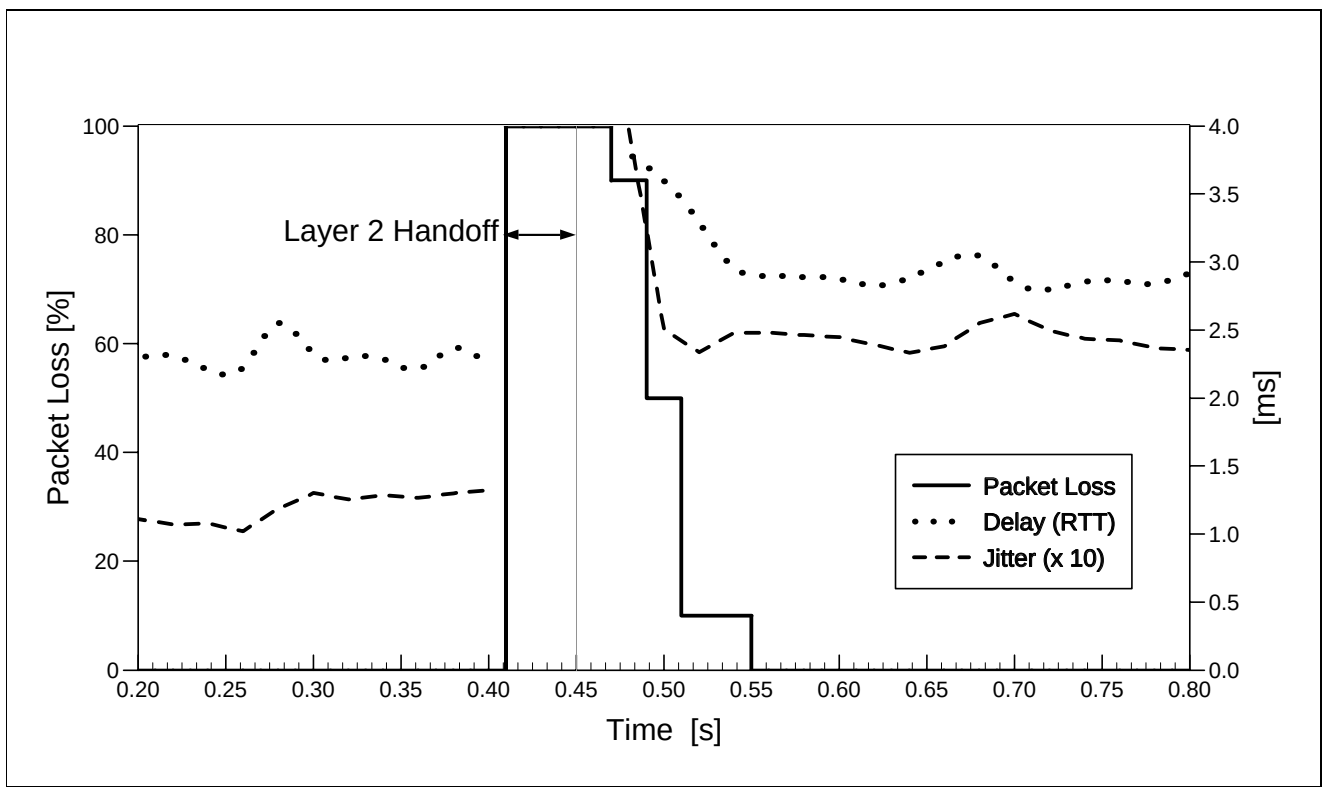

**Figure 2: Empirical Results on Router Advertised Handover**

The handover effect is dominated by packet loss, which is due to L2 handoff (40 - 50 ms) and MIPv6 address updates (10 – 60 ms), where binding update times take only a few milliseconds in our set-up. Duration of layer 2 transition we find in rough agreement with the report in [8]. The subsequent link local IP readdressing attained an average duration of 25 ms. Note that the Linux MIP stack discovers a change of network only through layer 3 router advertisements, which in our scenario, still conformal to [5], produce a base load of 0,5 % 802.11b network capacity.

[i] System details had been as follows: Debian Linux Version 2.4.19 with MIPL mobile IPv6 0.9.4, FreeBSD 4.6-STABLE with rtadvd, Win2000 with MSR (1.4) TCP-IPv6 driver 5.0.21955.1620.

It is a quite inefficient procedure to learn the arrival in a new network from router advertisements. The appropriate algorithm for a MN should be the discovery of the network change through the local L2 stack, instead, followed by an actively performed router solicitation call. Even though there is no standard for signaling 802.11 L2 events to IP, ignoring these to the MN locally available information cannot be justified. Like for wired NICs and their link statuses acquisition of layer 2 roaming event knowledge should be included in MIP stack implementations. Currently only the Windows MIP stack (MSR 1.4) is aware of L2 handoffs by means of Windows' proprietary architecture. Due to other prohibitive implementation problems of the stack it cannot be employed to profile relevant measurements on L2 triggered handovers.

Assuming L2 triggers in presence MIPv6 stateless local handovers may become impressively fast. A reduction in the timescale to the order of 1 - 3 milliseconds of the solicited router discovery handshake, the MIPv6 [5] timer variables MAX_RA_DELAY_TIME and MAX_RTR_SOLICITATION_DELAY at the router and the MN, may then reduce MIP local readdressing time well below 5 ms without adding base load to the network. For comparison note that the 802.11b clock tick allows for sending 10 packets of the size of a Router Advertisement per ms. Figure 3 visualises the theoretical steps and their temporal dimension.

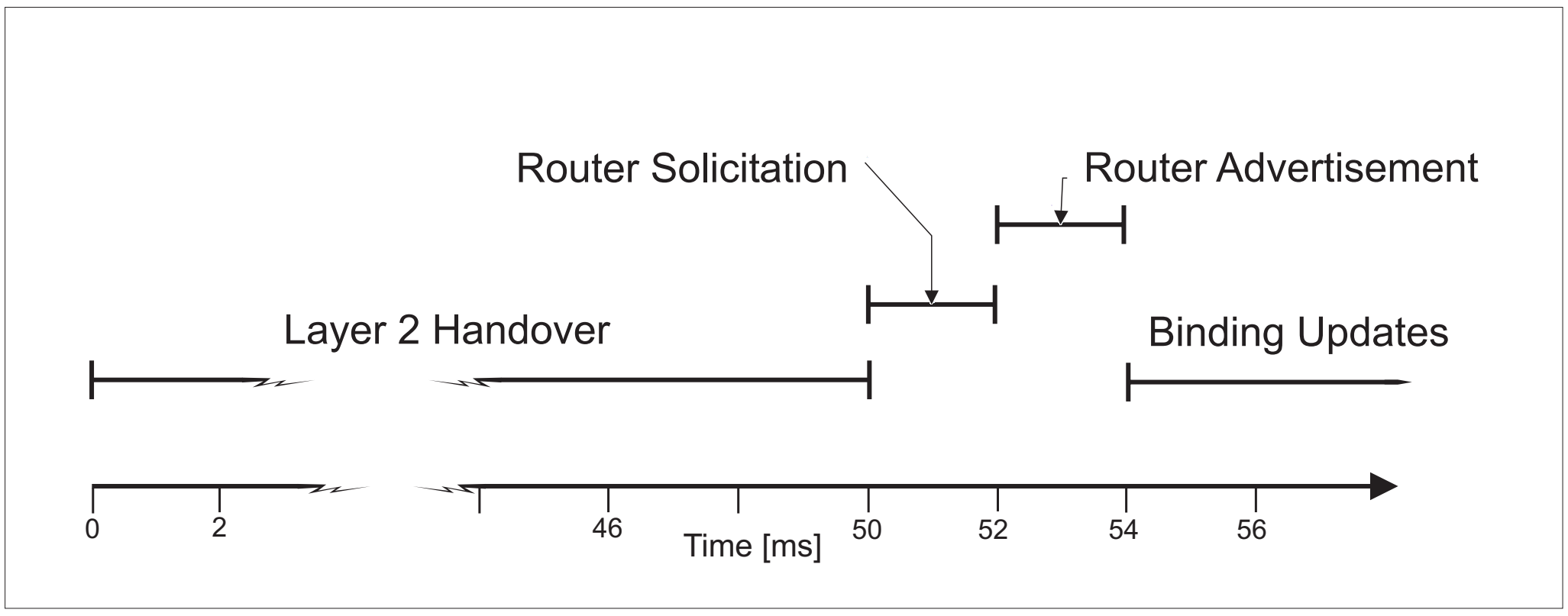

**Figure 3: Layer 2 Triggered Handoff over Time (Theory)**

# 4. Rethinking Real-Time Mobility

## 4.1. Topological Impacts

In the previous section it was shown that the period of local MIP handover procedures may be easily reduced to a marginal magnitude. A general topological scenario, though, may admit Home Agent and Correspondent Node at significant distance with respect to the Mobile Node, and Binding Updates (BU) cannot be neglected. Handoffs will partly disrupt, partly disturb continuous communication streams until BUs for HA and CN are completed. The total time for handover procedures may be written as

$$t_{handoff} = t_{local} + t_{BU-of-HA} + t_{BU-of-CN} \qquad (1).$$

Denoting by $t_A$ the roundtrip delay between Node A and the MN, suppressing a binding acknowledge from the CN [5] and assuming that the roundtrip delay between the CN and the HA can be approximated by $t_{CN}$ (s. Appendix A for a derivation), (1) transforms into

$$t_{handoff} \approx t_{local} + \tfrac{3}{2} t_{CN} + 2t_{HA} \qquad (2).$$

Here the coefficient preceding the delay times mainly originates from the return routability procedure. Following a similar approximation argument, the relative jitter enhancement at the time of communication re-establishment reads

$$\frac{Jitter_{handoff}}{Jitter_{stationary}} \approx \frac{t_{HA} + t_{CN}}{t_{CN}} \qquad (3).$$

Even though the coefficient in (2) may be reduced by optimizing secure updates with the help of cryptographically generated addresses [14] in the future, $t_{CN}$ and $t_{HA}$ may become quite large. In addition the relative magnitude of the mobile's distances to the HA and CN display crucial impact on Jitter enhancement, as we see from (3). It is therefore desirable to remove direct dependencies of $t_{CN}$ and $t_{HA}$ in equations (2) and (3). This can essentially be achieved with the help of two techniques: Constructing proxy components, on the one hand, and hiding update delays to the node communication on the other hand.

## 4.2. Topologically Robust Handover

Efficient, optimised handovers in MIPv6 are requested to remain within a real-time compliant temporal bound, independent of actual network geometry. It is also desirable to arrive at universal mobility mechanisms, supportable within a lean infrastructure. Central questions here rank around the presence of proxy agents.

Communicating end nodes cannot be substituted by proxies. Hence proxy techniques may reasonably be applied to the Home Agent, only. Such a concept for representing Home Agents in a distributed fashion has been developed within the Hierarchichal Mobile IPv6 (HMIPv6) [12]. While away from home, the MN registeres with a nearby Mobility Anchor Point (MAP) and passes all its traffic through it (s. figure 4). The vision of HMIPv6 presents MAPs as part of the regular routing infrastructure. The MN in the concept of HMIPv6 is equipped with a Regional Care-of Address (RCoA) local to the MAP in addition to its link-local address (LCoA). When corresponding to hosts on other links, the RCoA is used as MN's source address, thereby hiding local movements within a MAP domain. HMIPv6 reduces the number of 'visible' handover instances, but - once a MAP domain change occurs - binding update procedures need to be performed with the original HA and the CN. For this reason the HMIPv6 handover approach still directly depends on $t_{CN}$ and $t_{HA}$, even though the likelihood of visible handovers is reduced.

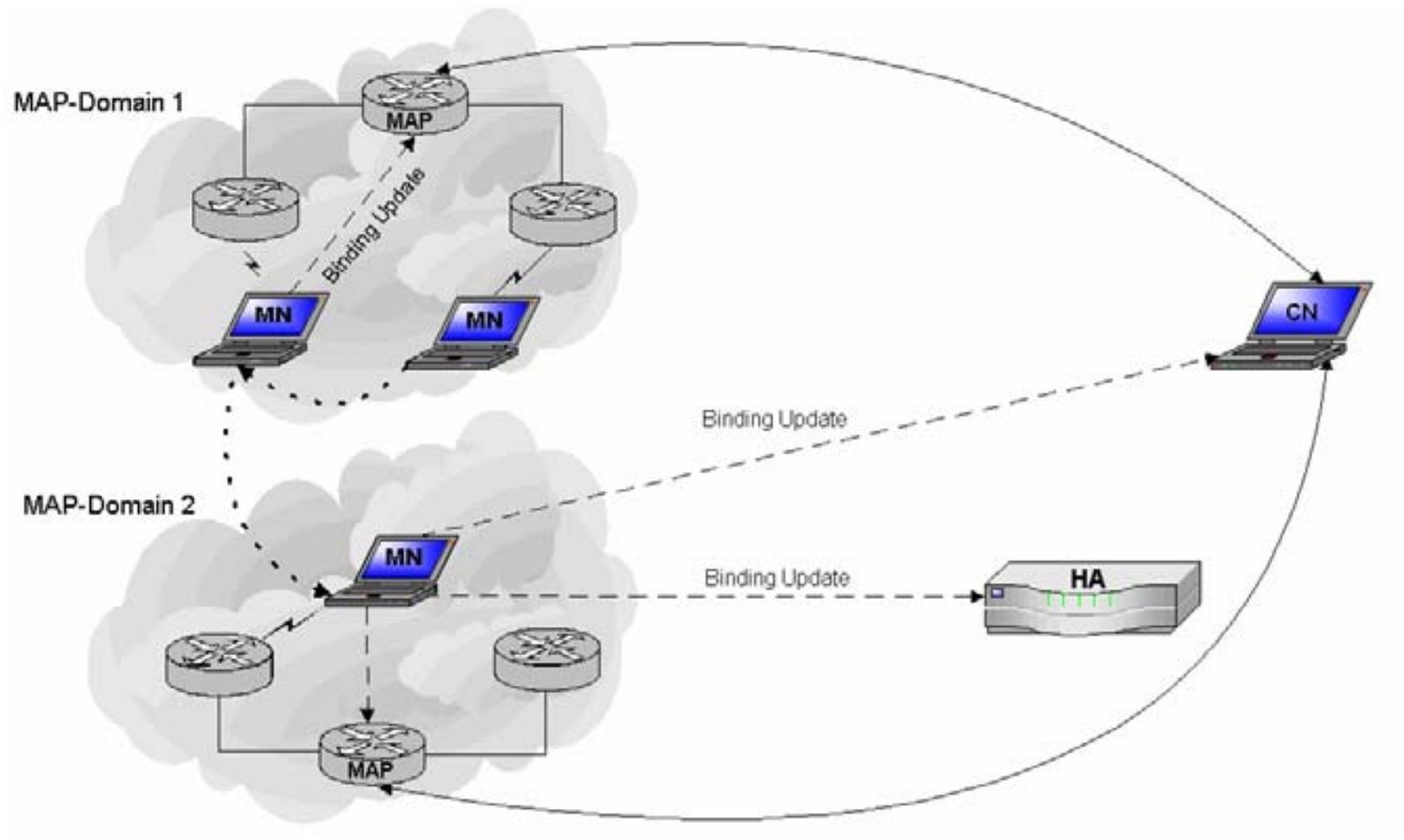

**Figure 4: Handover in a Hierarchical MIPv6 Environment**

The alternate approach adds handover hiding and is introduced in the Fast Handover for MIPv6 scheme (FMIPv6) [13]. FMIPv6 attempts to anticipate layer 3 handovers and to redirect traffic to the new location, as MN is about to move towards it. FMIPv6 extensively relies on layer 2 information and a layer 2 to 3 topology map, which is not present in current networks. Consequently this approach requests for layer 2 specific extensions.

FMIPv6 aims at hiding the entire handover process to communicating end nodes at the price of placing heavy burdens onto layer 2 intelligence. Severe functional deficits arise from incompleteness or tardiness of L2 negotiations and from the risk of a conceptual uncertainty: As the exact moment of layer 2 handover generally cannot be foreseen, and even flickering may occur, a traffic anticipating redirect may lead to data damage largely exceeding a regular MIPv6 handover without any optimization. Because of these unpredictable conditions, it remains a matter of accident for each handover, whether FMIPv6 will succeed in covering the handoff time and eliminate $t_{CN}$ and $t_{HA}$ dependence.

Regarding all the above ideas in the context of our discussion, some essential conclusions can be drawn:

1. Local proxy agents are needed to resolve the temporal dependence of handover procedures on the distances MN to CN and MN to HA.
2. Delay hiding is needed, as a new communication path of the MN cannot be established without updating the CN and the HA. The previous communication channel can be reliably used, while the new path is established.

A properly designed proxy architecture is presented by the HMIPv6 environmental approach: Sparsely distributed mobility agents, acting on a domain range, will minimize infrastructural effort and inter-proxy handovers. In contrast to the ‘greedy’ handover architecture of FMIPv6, the proxy architecture of HMIPv6 is additionally well suited to support multicast mobility, as discussed in the following section.

For hiding handover delay we suggest an approach quite opposite to FMIPv6: When a mobile in a HMIPv6 domain changes into a new network, it immediately should process a Binding Update with its previous MAP, aiming to preserve its formerly established communication channel. The previous MAP then should redirect its (tunnel-)forwarding to MN new LCoA. While processing its binding updates, a MN will thereby remain present in the network, it physically just left.

As has been already noted in [12], a MN can easily continue to receive packets through the previous MAP, after having performed a (vicinal) binding update. Reversely a MN is able to continue sending packets with its previous RCoA by using the previous MAP as a (tunnel-) forwarder.

However, to continue communication via the previously established path while the BU with the CN is processed, the MN is required to communicate with the CN using two addresses: Its previous address for forwarding and its new RCoA in the Binding Update. The Correspondent Node needs to allow such dual homing by preserving its previous Binding Cache entry, until the mobile has started to use its successfully established new RCoA. Dual Binding Cache entries are a minor extension to the MIPv6 draft [5].

In following this procedure of a shuffling reactive handover, a Mobile Node is enabled to rejoin communication after a local BU of MAPs, and perform a seamless IP handover, once distant binding updates have been fulfilled. In this way, no dependencies on distant delay times $t_{CN}$ and $t_{HA}$ remain. Note that this algorithm generalises to utilisation of the last successfully established communication channel / MAP in the case of MNs rapid movement.

## 4.3. The Tunnelling Problem

The packet forwarding to and from the Mobile Node is often organized by means of a tunnel section spanned to a static anchor component such as a MAP or a Home Agent. Through this technique a MN can hide its movement to CNs or to the routing infrastructure. Multicasting and the HMIPv6 [12] especially rely on tunnelling major traffic portions. However, keeping in mind real-time data requirements, it is highly desirable to avoid packet encapsulation. Besides the unwanted overhead, a tunnel may hide QoS information of the original packet headers and may require load and jitter generating packet fragmentation, if the tunnel entry point is distinguished from the sender.

Tunnelling can be avoided by a direct packet forwarding of the static anchor components. Such forwarding requires a change of packet’s source or destination address at the forwarder, which normally conflicts with checksums covering IPv6 pseudo headers. Most communication to and from a Mobile Node though carries MIPv6 extension headers, the type 2 routing header or the home address destination option header. These headers denote mobility independent destination or source addresses, which enter pseudo headers instead of original IPv6 header addresses.

All packets that carry a type 2 routing header, therefore can be forwarded by the MAP to the MN by changing the destination address, and all packets that carry a home address option can be forwarded from the MN by changing the source address, both without recalculation of header checksums. Packets arriving for the MN without routing header, e.g. packets sent to the home address, only, or multicast traffic, still need to be either tunnelled or rebuilt with a routing extension at the forwarding anchor point or Home Agent.

# 5. Multicast Mobility in a HMIPv6 Proxy Environment

## 5.1. Generals

Multicast-based packet distribution plays an important role in real-time applications, as it provides the only efficient, scalable solution for group communication. However, multicasting itself conceals complex mechanisms for group membership management and routing, which both are of slow convergence. In an IGMP/IPv4 environment, membership joins may take a period of the order of 30 s to become effective [7], which may significantly reduce in the presence of unsolicited joins, when a mobile enters a new network. Multicast routing essentially relies on source specific trees, constructed in a temporal dimension up to the order of 3 minutes, depending on the protocol in use. The use of shared trees as in PIM-SM or new routing algorithms may lead to a significant acceleration.

In general, the roles of multicast senders and receivers are quite distinct. While a client initiates a local multicast tree branch, the source forms the root of the entire tree. Hence multicast mobility at the sender side poses the more delicate problem. Multicast mobility is essentially left as an unfinished problem in MIPv6 [5]. It is worth recalling that multicast transport is stateless and unreliable. Packet loss, disorder or duplication has to be handled by application layer protocols s. a. RTP. It is also important to be aware that applications for multicast group communication normally are aware of source addresses, which must not change during ongoing sessions. Multicast mobility thus needs to provide mobile address transparency, as MIPv6 does in the unicast case. But at the same time multicast cannot follow MIPv6' strong communication relations, as they contradict the nature of speaking to groups.

Following the discussion of the previous section smooth handovers in MIPv6 cannot be assured without proxying home agents, as e.g. in HMIPv6 [12]. Keeping this perspective our subsequent proposition on multicast mobility embeds into the HMIPv6 unicast architecture, which routes all multicast traffic of a Mobile Node through a local MAP. No new protocol messages are needed for the proposed scheme. Any multicast traffic forwarding is transparent to routing infrastructure. So no dependencies on multicast routing protocols are imposed. For details of the corresponding M-HMIPv6 protocol see [17].

## 5.2. Mobile Multicast Receiver

Away from home the MN is registered with a local MAP and – beside its On-Link CoA (LCoA) – equipped with a Regional CoA (RCoA). To join a multicast group the MN tunnels its multicast group subscription, the MLD Listener Report [15], through its current MAP using RCoA as source address. The MAP will record the group address in its Binding Cache in order to

- subscribe for and preserve MNs multicast group membership,
- forward multicast packets to the Mobile Node.

Multicast packet forwarding here is identical to the unicast case (s. also section 4.3). Mobility within a MAP domain resultantly remains completely transparent to multicast routing procedures and does not impose any additional delay compared to the unicast case.

When changing MAP domains the MN simultaneously

- submits a Multicast Listener Report within the new MAP domain and
- sends a Binding Update with its new LCoA to the previous MAP

in addition to its regular HMIPv6 handover signalling.

On receiving MNs Binding Update the previous MAP forwards multicast packets to the foreign domain. This Binding Update should either contain an appropriately short lifetime or, on reception of the multicast stream through the new MAP, the MN should send a Binding Update with zero lifetime to the previous MAP to eliminate its Binding Cache entry and end packet forwarding.

Note that the MN should preserve the previous multicast MAP as a forwarder until a new multicast branch is established. In the case of rapid movement this may lead to an old anchor point persisting through several hops. There may be a short time of duplicate packet reception at the MN. By applying common vicinity arguments it can be assumed that handover disruptions of multicast streams are even less significant than in the unicast case.

### 5.3. Mobile Multicast Source

A MN in a foreign MAP domain initiates multicast-based communication by sending packets through its MAP, using RCoA as its packet source address. A multicast routing source tree originating at the MAP may thus be constructed, if requested by the routing protocol in use. Local movement within a MAP domain remains transparent to multicast routing. As receivers are aware of source addresses and as the mobile RCoA address may change, a Home Address Destination Option must be included. Note that the presence of Home Address Option deviates from MIP [5]. Furthermore, a multicast listener receiving packets with Home Address Option, in general need not have a Binding Cache Entry for the correspondent Home Address. A CN therefore cannot verify multicast packets with respect to its Binding Cache. This processing of multicast packets at the receiver extends the MIPv6 unicast scheme.

Upon arrival in a foreign MAP domain the MN continues its multicast transmission via the previous MAP, using its previous RcoA, and simultaneously starts sending through its current MAP (s. figure 5), thereby initiating a new multicast routing tree with the new RCoA as source address. Packets may be bi-casted or, alternatively, sent as sparsely initiated empty probes. The idea, of course, lies in performing the MAP handover as soon as the new multicast tree is completed. However, there is no meaningful inquiry on the establishment of multicast routing trees (a MN listening into the group, e.g. via the previous MAP or HA, results in strongly topology dependent information).

A Mobile Node should stop sending its multicast streams through the old MAP after a reasonable timeout is reached. The preconfigured value for this timeout should be a quantity characteristic to protocol convergence. It may be derived from timeout values of the multicast routing protocol employed within the infrastructure. Note that this value may reach up to the order of minutes [17]. If a MN leaves home, the HA attains the role of the previous MAP.

In case of rapid movement, the MN needs to stay with its formerly established (or first) MAP. When experiencing a domain handover prior to reaching its handover timeout, a MN terminates multicast traffic to the intermediate MAP, initiates communication through the current MAP and resets its timer, while continuously sending through the formerly old MAP.

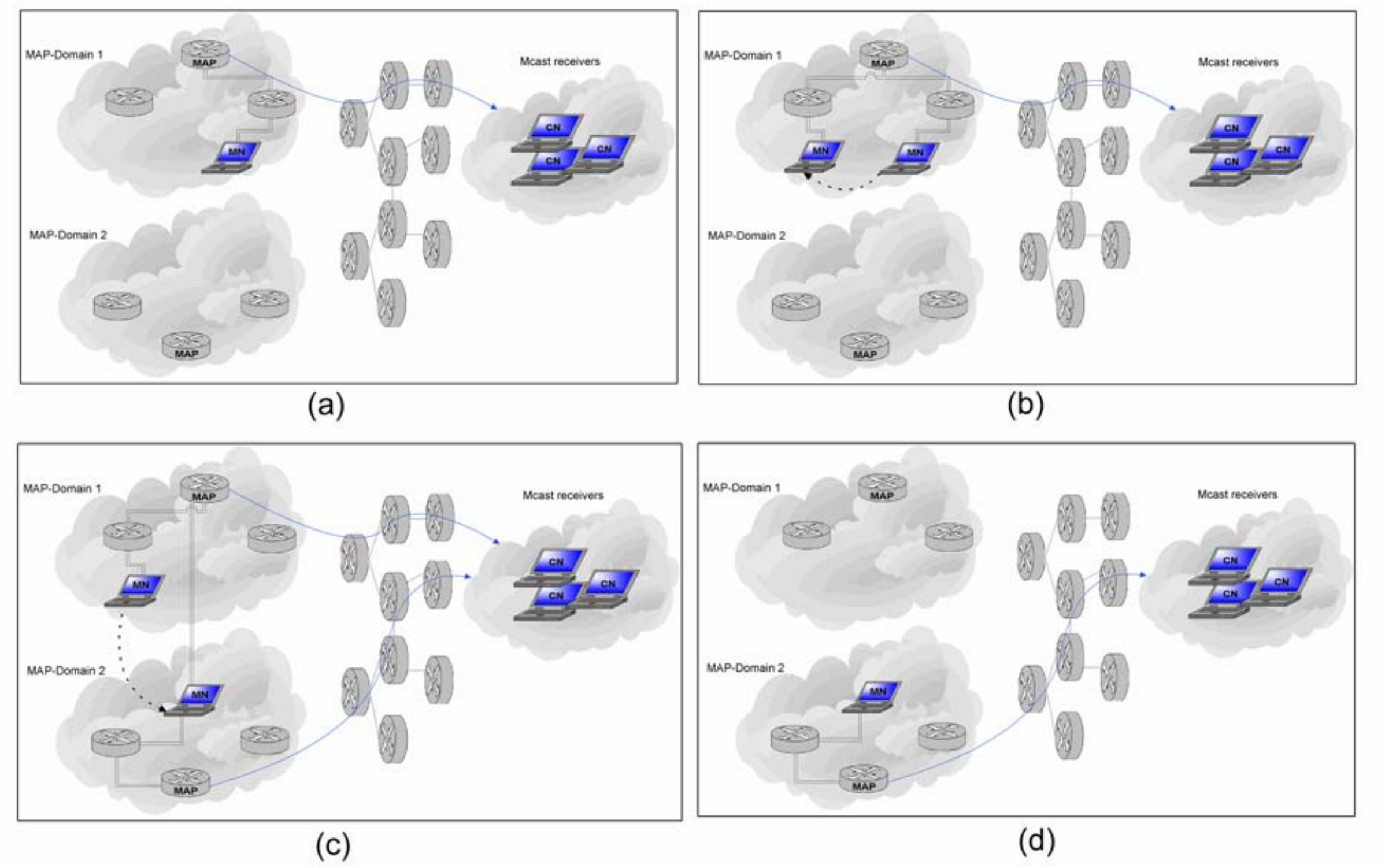

**Figure 5: Roaming of Multicast Senders in a HMIP Environment. The Multicast Source Tree originating at the local MAP (a) remains unaffected by local Movements (b). When changing a MAP domain the MN continues to send through the previously established Routes (c) while initiating a new Tree from its current MAP. Finally the previous Path is diminished (d).**

The procedure proposed above aims at minimizing disturbances or disruptions of multicast communication originating from a mobile sender at the price of network efficiency during handover periods. It is likely that group members receive duplicate streams during some portion of handover time, even though redundant traffic can be minimized by following the tree initialization procedure described in [17].

To judge the overall feasibility of the method, resource load at the MN / network as well as application compliance with bi-casted data streams need consideration.

However, as multicast routing is expected to improve intervals of overlapping traffic will decrease to a more feasible timescale.

## 6. Conclusions and Outlook

Mobility is one of the most challenging and demanded developments in IP networks today. Starting from real-time requirements we analyzed the temporal handoff performance of the current draft statuses in the present paper. Continuing the discussion along common approaches for handover acceleration we presented improvements for a seamless, efficient mobility support in a hierarchical Mobile IPv6 environment. The proposed communication scheme under handover no longer depends on the roundtrip delays for updates of bindings with Home Agents or Correspondent Nodes.

Extending the unicast concepts we introduced a treatment for mobile multicast clients *and* sources. To empirically analyze and examine the introduced protocol mechanisms we will concentrate on empirical analysis and simulations of the proposed solutions in future work.

## Appendix A

We want to derive the approximations (2) and (3) of section 4.1 in this appendix. Starting from

$$t_{handoff} = t_{local} + t_{BU-of-HA} + t_{BU-of-CN}\,, \qquad \text{(A.1)}$$

and neglecting effects of stack details, we want to evaluate the consecutive Binding Update periods with HA and CN. Now clearly

$$t_{BU-of-HA} \approx t_{HA}\,, \qquad \text{(A.2)}$$

if $t_{HA}$ denotes the roundtrip delay between the MN and the HA.

To approximate the BU with the CN, the return routability procedure needs to be evaluated. Keeping the notation of $t_A$ for the roundtrip delay between Node A and the MN, $t_{HA-CN}$ for the roundtrip delay between HA and CN, we arrive at

$$t_{BU-of-CN} \approx \tfrac{1}{2}\left\{\max(t_{CN}, t_{HA-CN} + t_{HA}) + \max(t_{CN}, t_{HA-CN} + t_{HA}) + t_{CN}\right\}, \qquad \text{(A.3)}$$

where right-hand terms represent (in order) Test Init messages, Test messages and BU. Under the perspective that the MN and CN are located arbitrarily, but not close to the HA, we may approximate $t_{HA-CN}$ by $t_{CN}$. Then (A.3) reads

$$t_{BU-of-CN} \approx \tfrac{3}{2} t_{CN} + t_{HA}\,, \qquad \text{(A.4)}$$

and equation (2) in section 4.1 follows by substituting (A.4) and (A.2) into (A.1).

To derive the interarrival jitter rate approximation, we make the simplifying assumption of a homogeneously distributed jitter generation along the network lines. Thus jitter is proportional to network dimensions. Furthermore we assume that $t_{CN}$ does not significantly change under handover and, as for (A.4), may approximate $t_{HA-CN}$.

At the time of communication re-establishment packets arrive by triangular routing via the HA. Thus the jitter after handoff becomes proportional to $t_{HA-CN} + t_{HA}$ as compared to $t_{CN}$ before handoff. Applying approximations the rate equation (3) of section 4.1 follows directly.

## Acknowledgement:

This work was supported in part by the German Bundesministerium für Bildung und Forschung.

We would like to thank Stefan Zech and Mark Palkow for their cheerful collaboration: They not only suffered through many clarifying discussion, but also provided valuable insight into numerous relevant details.

## Vitae:

**Thomas Schmidt** is the head of the computer centre of FHTW Berlin and appointed professor of Information Engineering at HAW Hamburg. He studied mathematics and physics at Freie Universität Berlin and University of Maryland, USA. In 1993 he received his PhD in mathematical physics for a work in many particle theory of quantum mechanics done at the theory group of the Hahn-Meitner-Institut Berlin. Since the late 1980s he has been involved in many computing projects, especially focusing on simulation and parallel programming, distributed information systems and visualisation. His current fields of interest lie in the areas of multimedia networking and hypermedia information processing, where he has continuously conducted numerous projects on national and international level.

**Matthias Wählisch** is a member of the networking group of the computer centre of FHTW Berlin. He is studying mathematics and computer science at Freie Universität Berlin. His major fields of interest lie in networking protocols where he looks back on five years of professional experience in project work and publication.